\documentclass[12pt, preprint]{aastex}
\usepackage{graphicx}
\bibstyle{apj}
\begin{document}

\title{Suzaku Observations of the Black Hole H1743-322 in Outburst}

\author{J.L.Blum\altaffilmark{1}, J.M. Miller\altaffilmark{1}, E. Cackett\altaffilmark{1,7},
K. Yamaoka\altaffilmark{2}, H. Takahashi\altaffilmark{3},
J. Raymond\altaffilmark{4}, C. S. Reynolds\altaffilmark{5}, A. C. Fabian\altaffilmark{6}}

\altaffiltext{1}{Department of Astronomy, University of Michigan, 500 Church Street, Ann Arbor, MI 48109}
\altaffiltext{2}{Department of Physics and Mathematics, Aoyama Gakuin University, Shibuya, Tokyo, Japan}
\altaffiltext{3}{Department of Physical Science, Hiroshima University, 1-3-1 Kagamiyama, Higashi-Hiroshima,
                \indent{\indent{Hiroshima, 739-8526 Japan}}}
\altaffiltext{4}{Harvard-Smithsonian Center for Astrophysics, 60 Garden Street, Cambridge, MA 02138}
\altaffiltext{5}{Department of Astronomy, University of Maryland, College Park, MD 20742}
\altaffiltext{6}{Cambridge University, Institute of Astronomy, Madingley Road, Cambridge CB3 0HA, UK}
\altaffiltext{7}{Chandra Fellow}

\begin{abstract}
We observed the Galactic black hole candidate H1743-322 with \emph{Suzaku} for
approximately 32 ksec, while the source was in a low/hard state during its 2008 outburst. ÊWe collected and analyzed the data with the HXD/PIN, HXD/GSO and XIS cameras spanning the energy range from 0.7$-$200 keV.  Fits to the spectra with simple models fail to detect narrow Fe XXV and Fe XXVI absorption lines, with 90\% confidence upper limits of 3.5 eV and 2.5 eV on the equivalent width, respectively.  These limits are commensurate with those in the very high state, but are well below the equivalent widths of lines detected in the high/soft state, \emph{suggesting that disk winds are partially state-dependent}.  We discuss these results in the context of previous detections of ionized Fe absorption lines in H1743-322 and connections to winds and jets in accreting systems.  Additionally, we report the possible detection of disk reflection features, including an Fe K emission line.

\end{abstract}

\section{Introduction}

H1743-322 was discovered during its 1977-1978 outburst with the
\emph{High Energy Astronomical Observatory-1 (HEAO-1)} and \emph{Ariel
5} (Doxsey et al. 1977, Kaluzienski \& Holt 1977).  Based on its spectral characteristics,
H1743-322 was classified as a Galactic X-ray transient by
White \& Marshall (1984).  The 2003 outburst of H1743-322 allowed for two
discoveries that provided evidence that this source harbors a black
hole primary.  First, Homan et al. (2003) reported the detection of
quasi-periodic oscillations (QPOs) at 160 Hz and 240 Hz, a 2:3
frequency ratio that is typical for black hole candidates (e.g. GRO
J1655-40: Strohmayer 2001a; GRS 1915+105: Strohmayer 2001b; XTE
J1550-564: Miller et al. 2001).  Second, Rupen et al. (2003) (see also
Corbel et al. 2005) reported a relativistic jet (v/c$\simeq$0.8) in
H1743-322, which allows this source to be termed a ``microquasar.''

Disk-jet coupling, i.e., the relation between inflow and outflow, has become one of the richest areas
of X-ray binary research.  The ``low/hard'' and the ``high/soft'' states are the two most 
diametrically opposed in regards to jet formation.  The low/hard state is characterized by an energy spectrum
that is dominated by a power-law-like component.  A photon index of $\sim$ 1.6 and a high-energy cutoff at $\sim$ 100 keV (Fender \& Belloni 2004) are typical
for a black hole X-ray binary in the low/hard state.  A weak soft X-ray component, most likely associated with a thermal disk (e.g. Miller et al. 2006a, 
Reis et al. 2009), can also be present.  In contrast, a strong thermal component, which can be modeled as a
disk-blackbody with an inner temperature of 1$-$2 keV, is prevalent in
the high/soft state (Fender \& Belloni 2004).  When both GRS 1915+105 (Dhawan et al. 2000) and Cyg
X-1 (Stirling et al. 2001) were in the low/hard state, VLBI radio images have
shown a spatially-resolved radio jet during periods of quasi-steady
radio and hard X-ray emission.  However, in the high/soft state there
is a dramatic drop in the radio emission compared to the low/hard
state (e.g. Fender \& Belloni 2004).  Although the nature of radio emission
in the ``very high state'' remains unclear, McClintock \& Remillard (2006) suggest that
this state is relatively radio quiet.  To date, no source has exhibited optically-thick radio emission consistent with a jet in the high/soft state.

In this paper, we refer to observations of H1743-322 in the high/soft state and ``very high state'' made in 2003 (Miller et al. 2006c).  ``Intermediate'' and  ``very high'' (or steep power-law) states appear to be quasi-steady states that share some of the characteristics of 
both the low/hard and high/soft states.  The energy
spectrum possesses strong thermal \emph{and} power-law components with
no apparent high-energy cutoff.  The photon index is typically $\geq$
2.4, which is steeper than the index seen in the low/hard state
(McClintock \& Remillard 2006).  Transitions between the high/soft and low/hard states frequently pass through intervals
of these transitional states.  While some suggest that the very high state is in itself an ``intermediate'' state, we acknowledge the very high state as a bona fide state of black hole binaries in accordance with McClintock \& Remillard (2006).    

A disk wind can be key to understanding the nature of the accretion
flow in disk systems (e.g. Proga 2003, Ueda et al. 2009,
Ohsuga et al. 2009).  Previously, \emph{Chandra} observations of
H1743-322 during its 2003 outburst found evidence of an ionized disk
wind in the high/soft state, in the form of blue-shifted Fe XXV and
XXVI absorption lines (Miller et al. 2006c).  Figure 1 compares the light curves of H1743-322 when it was observed in the high/soft state and the low/hard state.  The apparent difference in counts/sec and hardness between the two states emphasizes the distinction between spectral states for black hole binaries. Iron aborption lines were absent when
\emph{Chandra} observed H1743-322 in the very high state.  Similar
results were obtained from \emph{Chandra} observations of GRO J1655-40
(Miller et al. 2006b, Miller et al. 2008) and GRS 1915+105
(Miller et al. 2008, Neilson \& Lee 2009).  The physical mechanism needed
to produce such a disk wind is still being investigated, aided by
an extensive theoretical framework (see e.g. Blandford \& Payne 1982,
Begelman et al. 1983, Proga et al. 2000).  In this paper, we report the
results of an analysis of the \emph{Suzaku} X-ray spectra resulting
from the H1743-322 October 2008 outburst (see Figure 4).  Preliminary
results from this observation were reported in Blum et al. 2008 (ATel
\# 1841).

\section{Data Reduction}
\emph{Suzaku} observed H1743-322 on 2008 October 7 starting at 16:19:21 (UT).  The observation 
duration was approximately 72 ks.  The X-ray Imaging Spectrometer (XIS) pointing position was used.  The three detectors constituting the X-ray Imaging Spectrometer
(XIS0, XIS1, XIS3; energy range 0.2$-$12 keV) were each operated in 3x3 and 5x5 editing modes; using the 1/4 window mode with a 1.0 sec burst option helped to limit photon pileup.    No timing mode was used.  The energy resolution (FWHM) of the XIS cameras is $\sim$ 120 eV at 6.0 keV and $\sim$ 50.0 eV at 1.0 keV.  Each CCD camera has a single CCD chip with an array of 1024 x 1024 pixels, and covers an 18$\arcmin$ x 18$\arcmin$ region on the sky.  Each pixel is 24$\mu$m square, and the size of the CCD is 25mm x 25mm.  The XIS1 sensor uses a back-side illuminated CCD, while the other two use a front-side illuminated CCD (e.g. Koyama et al. 2007).  XIS on-source times of approximately 32 ks were achieved.  This resulted in a dead-time corrected net exposure of 16 ks for the XIS cameras.  

Soft X-ray spectra within the energy ranges 0.7$-$1.5 keV and 2.2$-$10.0 keV were used. The 1.5$-$2.2 keV energy range was ignored to avoid systematic errors due to calibration in the Si band.  We used the latest calibration database available at the time of analysis (CALDB 20090925).  From the cleaned event files we extracted the source
light curve and spectrum with \texttt{xselect}.  A circle
centered on the source with a radius of 129$\arcsec$ (123 in pixel units) was used for the source extraction region.  A circular, off-source region with a radius of 45$\arcsec$
(43 in pixel units) was used to extract the background. The 3x3 and 5x5 mode event files for each XIS detector were loaded together into \texttt{xselect}.  A single source and background spectrum was extracted from these event files for each XIS detector.
The XIS redistribution matrix files (RMFs) and ancillary response files
(ARFs) were created using the tools \texttt{xisrmfgen} and \texttt{xissimarfgen}
available in the HEASOFT version 6.6.2 data reduction package.  The 3x3 and 5x5 mode event files were used
to specify good time intervals and the data was grouped
to a minimum of 300 counts per bin using the FTOOL
\texttt{grppha}.  

The Hard X-ray Detector (HXD) covers the energy range 10.0$-$700 keV and is a combination of Positive Intrinsic Negative (PIN) silicon diodes and Gadolinium Silicate (GSO) phoswich counters.  A net exposure of 29 ks was achieved using the HXD (PIN and GSO) cameras, which were operated in their normal mode.  The HXD features an effective area of $\sim$ 160 $\mathrm{cm^{2}}$ at 20.0 keV, and $\sim$ 260 $\mathrm{cm^{2}}$ at 100 keV (e.g. Takahashi et al. 2007).  The HXD time resolution is 61 $\mu$s.  The HXD-PIN/GSO data were reprocessed from an unscreened event file with up-to-date calibration databases in accordance with the \emph{Suzaku} Data Reduction Guide (CALDB 20090511).  The appropriate response matrices for the XIS aimpoint and background files were downloaded from the \emph{Suzaku} web page.  In our analysis, the energy range for the HXD cameras was restricted to 12.0$-$200 keV.  

\section{Data Analysis and Results}

Using the X-ray spectral fitting software package (XSPEC v. 11.3.2,
Arnaud 1996), the \emph{Suzaku} H1743-322 spectra from the three XIS
detectors and the HXD detector were fit jointly.  A normalizing
constant (ranging between 1.0 and 1.1) was allowed to float between spectra to account for different
detector flux zero-points.  Our initial spectral model consisted of an
absorbed power-law ($\chi^{2}/\nu$ = 3348/3095).  We used
\texttt{tbabs} in our model to account for photoelectric
absorption in the interstellar medium (ISM).  Replacing a
simple power-law with a broken power-law provided an improved fit
($\chi^{2}/\nu$ = 3053/3093) with $\mathrm{E_{break}}$ $\simeq$ 36(2)
keV, $\Gamma_{1}$ $\simeq$ 1.55(1), and $\Gamma_{2}$ $\simeq$ 2.2(1).

The fit results in the 0.7$-$200 keV band with the absorbed
broken power-law model are reported in Table 1 with a spectral fit
being shown in Figure 2 and Figure 3.  For Figure 3 we note that the residuals near 10.0 keV in the XIS spectra are likely due to calibration uncertainties and not photon pile-up (different size extraction regions
had negligible effect on the resulting spectral fit).  The spectral parameters obtained
through our fits for H1743-322 are typical of a Galactic black hole
candidate in the low/hard state (McClintock \& Remillard 2006, Fender \& Belloni 2004).  
Note that the high equivalent neutral hydrogen column density (2.00(1) x $10^{22}$ $\mathrm{cm^{-2}}$) 
along the line of sight is consistent with a central Galactic distance to H1743-322 and is in accordance with the value
used by Miller et al. (2006c).  
The unabsorbed flux inferred in the 0.5-200 keV band is
(7.51 $\pm$ 0.08) x $10^{-9}$ erg $\mathrm{cm^{-2} s^{-1}}$.  Currently, the mass of
H1743-322 and the distance to the source are unknown.  But, given its
position near the Galactic center (\emph{l} $\simeq$
$357.255\,^{\circ}$, \emph{b} $\simeq$ $-1.833\,^{\circ}$) and
relatively high column density, it is reasonable to use a Galactic
center distance of 8.5 kpc.  Using this value as an estimate of the
distance to the source gives a luminosity of $\mathrm{L_{X}}$ =
(6.37 $\pm$ 0.07) x $10^{37}$  $(d/8.5 \mathrm{kpc})^{2}$ erg $\mathrm{s^{-1}}$ (we calculated the Galactic
ridge X-ray emission contribution to be negligible and accounted for
by our background spectrum).  This calculated luminosity indicates
that the source would be 0.08 $\mathrm{L_{EDD}}$ for a 6 $M_{\sun}$
black hole and 0.05 $\mathrm{L_{EDD}}$ for a 10 $M_{\sun}$ black hole.

A broken power-law may be a phenomenological description of Compton
upscattering of disk photons by a central, hot, electron-dominated
``corona'' (see e.g. Ross et al. 1999).  Therefore, we also fit a Comptonization
model (\texttt{compTT}; Titarchuk 1994) to the broadband \emph{Suzaku} data.  This is an analytic model
describing Comptonization of soft photons in a hot plasma. The approximations used in the model 
work well for both the optically thin and thick regimes.  Fitting the \texttt{compTT} model resulted in an electron temperature
of kT $\simeq$ 38(5) keV for a disk temperature of kT $\simeq$ 0.25(2) keV (typical disk
temperature for the low/hard state; see e.g. Miller et al. 2006a).  The
reduced $\chi^{2}$ for this Comptonization model was $\simeq$
3066/3092 (see Table 2).  Note that the energy break in our best-fit broken
power-law model, which can indicate the electron temperature of the
purported corona, is consistent with the electron temperature in the
Comptonization model.  
 
To test for the presence of the X-ray Fe XXV and Fe XXVI absorption lines in the
\emph{Suzaku} spectrum for H1743-322, we added a Gaussian component to
the broken power-law continuum model.  We froze the line energy at the theoretical values for Fe XXV (6.70 keV) and Fe XXVI (6.97 keV), respectively.  The line width was frozen at zero
(due to the likelihood that the lines would be at least as narrow as
those observed with \emph{Chandra}; see Miller et al. 2006c) while the line
flux normalization parameter (in units of photons $\mathrm{cm^{-2}
s^{-1}}$) was allowed to vary.  Error analysis was performed on the line flux
(using the \texttt{error} command) to derive the maximum allowable
iron line normalization at 90\% confidence.  The line wavelengths,
fluxes, and equivalent widths are listed in Table 3.  The equivalent width upper limits for both iron absorption lines were $<$ 3.5 eV. 

Based on these results, we conclude that these iron absorption lines were not present in the \emph{Suzaku} spectrum for H1743-322 when it was observed in the low/hard state.  The Miller et al. (2006c) \emph{Chandra} observations spanned several black hole binary spectral states, particularly the very high state and the high/soft state.  Similar to our \emph{Suzaku} results, the Fe XXV and Fe XXVI absorption lines were absent in the source's spectrum in the very high state (equivalent width upper limits $<$ 4.3 eV; see Miller et al. 2006c).  For softer states (i.e. the high/soft state), however, each detection of Fe XXV and Fe XXVI absorption lines was at the 4 $\sigma$ level of confidence or higher (with e.g. equivalent widths of 16(1) eV and 25(3) eV, respectively; see Miller et al. 2006c: Table 3).    The combined results of \emph{Suzaku} and \emph{Chandra}  indicate that there is a spectral state dependence on absorption lines, and therefore disk winds.  

Disk reflection features are common amongst Galactic black
hole X-ray spectra (Miller 2007).  As shown in Figure 4, there
is a potential Fe K$\alpha$ emission line present in the \emph{Suzaku}
spectra.  A Gaussian component was added to the broken power-law model
to get estimates on the strength of possible Fe K$\alpha$ emission
lines.  The 90\% confidence level on the strength
of narrow and/or broad Fe K$\alpha$ emission lines are given in Table 4
for line energy and full width at half maximum (FWHM) values.  The line energy and width were allowed to vary
from 6$-$7 keV and 0$-$3.0 keV, respectively, for the low ionization iron parameter.   This resulted in a low equivalent width of 17(6) eV and a line flux of
0.3(1) ($10^{-5}$) photons $\mathrm{cm^{-2}}$ $\mathrm{s^{-1}}$ with
FWHM = 0.6(3) keV.  A weak Fe K emission line in the
\emph{Suzaku} spectra (see Figure 3) of H1743-322 is statistically
significant, though the line has an equivalent width of just 17(6)
eV (F-value $>$ 8.0, making the line significant at more than the 4.0 $\sigma$ level of confidence).  Note that for low ionization iron (Fe I $-$ Fe XVI), a theoretical line
energy of 6.40 keV can reasonably be taken to probe Fe I - Fe XVI since the line
energy changes only 0.03 keV between those charge states (Kallman et al. 2004).  Also note that we made broadband reflection fits to the spectra.  For consistency, we investigated models for 
a relativistic line profile around a rotating black hole convolved with a broken power-law reflected from ionized matter (specifically, \texttt{tbabs x (laor +kdblur x bexriv)} ;  see Laor 1991, Magdziarz \& Zdziarski 1995).  However, 
these models were poorly constrained and were not statistically superior to the broken power-law and thermal Comptonization models. 

\section{Discussion}

We have analyzed a sensitive {\it Suzaku} observation of the Galactic
black hole candidate H1743-322 in the low/hard state.  The
continuum spectrum can be described using a broken power-law or thermal
Comptonization.  The spectrum is interesting for lacking evidence 
of an X-ray disk wind.  When combined with prior results, our findings suggest that disk winds are partially state-dependent.  There is also 
weak evidence for a neutral Fe K emission line that likely arises in the disk.

Previous observations show that disk winds in stellar-mass black holes
seem to be stronger in soft disk-dominated (high/soft) states than in
other states (e.g. Miller et al. 2006c, Miller et al. 2008,
Neilson \& Lee 2009).  Indeed, it is possible that winds and jets are
anti-correlated (Miller et al. 2006c, Miller et al. 2008, Neilson \& Lee 2009)
-- the disk may alternate its outflow mode.  In GRS 1915+105,
Miller et al. (2008) and Neilson \& Lee (2009) found broad He-like Fe XXV emission lines in the low/hard
state and highly ionized narrow H-like Fe XXVI lines in the high/soft
state.  Equivalent widths ranged between $\sim$ 5$-$30 eV for the
H-like Fe XXVI absorption lines (Neilson \& Lee 2009: Table 1), which are
consistent with those observed in the high/soft state \emph{Chandra}
observation for H1743-322 (Miller et al. 2006c).  Ueda et al. (2009) also detected both H-like Fe XXVI  and He-like Fe XXV absorption lines (with equivalent widths of $\sim$ 40 eV) in the high/soft state for GRS 1915+105.  While both sources differ
in subtle ways from the ``canonical'' spectral states for X-ray
binaries (GRS 1915+105: see Belloni et al. 2000; H1743-322: see
Homan et al. 2005), the prevalence of disk
winds in the high/soft state is clear.

Miller et al. (2006c) detected absorption lines when \emph{Chandra} observed
H1743-322 when it was in the high/soft state (see Homan et al. 2005,
Miller et al. 2006c).  The Fe XXVI line had a velocity width (FWHM) of $1900 \pm 500$ $\mathrm{km}$ $\mathrm{s}^{-1}$, suggesting a highly ionized medium in outflow.     The equivalent widths observed in the high/soft state for Fe XXV and Fe XXVI absorption 
lines were greater than those we observe in the low/hard state by a factor of approximately 4.4 and 10, respectively.  
However, those lines were absent in the very high state, with
tight upper-limits on the line flux.  Based on these observations, it
was not certain whether the wind was quenched in the very high state,
or if the ionization parameter in the wind had merely increased and
hindered its detection.  In the low/hard state we observed, the
9.0$-$20.0 keV ionizing flux is a factor of 2.3 lower than in the very
high state (Miller et al. 2006c), yet ionized Fe XXV and Fe XXVI absorption
lines are ruled out with more stringent upper limits.  This demonstrates
that disk winds are indeed state-dependent: strongest in the
disk-dominated high/soft state, and weak or absent in other states.  Furthermore, the lack of a disk wind in the hard state for the 2008 outburst and the detection of relativistic jets in the hard state during the 2003 outburst for H1743-322 (Corbel et al. 2005) supports the notion the disk winds and jets are anti-correlated.

Disk reflection features are common in a majority of Galactic black
holes' spectra in the low/hard state (e.g. Rossi et al. 2005,
Miller et al. 2006a).  Our analysis of H1743-322 indicates the
presence of such features, most notably, the Fe K$\alpha$ emission
line.  The predicted equivalent widths of the Fe K emission line
depend on the geometry, inclination, and ionization state of the
system.  For a centrally illuminated neutral disk with varying
power-law photon index (1.3 $\leq$ $\Gamma$ $\leq$ 2.3), an equivalent
width between 100 eV and 150 eV is expected for an inclination range
of 45$-$60 degrees (George \& Fabian 1991: Figure 14) .  For an ionized disk
(1.0 $<$ log $\xi$ $<$ 5.0, with $\xi$ being the ionization
parameter), equivalent widths can range between 100 eV and 500 eV
(Ballantyne et al. 2002: Figure 3).   Note that
Miller et al. (2006c) detected no reflection features for this source in
the high/soft state and the very high state.  Based on the calculations in George \& Fabian (1991), a possible explanation 
for our low equivalent width values is a
high inclination (i.e. $\theta$ $>$ $80\,^{\circ}$).  Viewing the
central engine through a corona seen at high inclination can serve to
diminish the clarity of reflection features (Petrucci et al. 2001).  Dips seen in the
 light curves for H1743-322 also indicate that it is likely viewed at a high inclination (Miller et al. 2006c: Figure 5).  

Additionally, by using the width (i.e. FWHM) of the Fe K emission line we can estimate how far the inner disk extends in H1743-322.  Note that there is high uncertainty in FWHM (0.6(3) keV), which only allows for lower and upper boundaries of the inner disk radius.  Based on our model (i.e. broken power-law added to a Gaussian component) and given in units of gravitational radii ($r_{g} = GM_{\mathrm{BH}}/c^{2}$), the inner disk ranges from $\sim$ 40$-$600$r_{g}$.  The innermost stable circular orbit (ISCO) around a nonspinning black hole is $r_{\mathrm{ISCO}}$ = 6$r_{g}$.  A possible explanation for the large differences between our estimates and the expected inner disk radius for a nonspinning black hole is that the inner disk does not extend to the ISCO and is truncated (see e.g. Esin et al. 1997).  However, given that our estimates are inferred from parameter values that are poorly constrained, we must regard our inferred radii with caution.  It is possible, for instance, that the detection of a red wing in a relativistic line is prevented by calibration uncertainties in the 2.0-3.0 keV band that can have a small impact on the continuum.  

Although examining disk winds and disk reflection in stellar-mass
black holes can be observationally challenging, the possibility of a
state-dependent anti-correlation between winds and jets may offer
insights into different AGN modes (i.e. radio-loud vs. radio-quiet).
\emph{Chandra}, \emph{XMM-Newton}, and \emph{Suzaku} can all make
important contributions in this regard.  In the near future, {\it
Astro-H} and {\it The International X-ray Observatory} will be able to
study winds and reflection in unprecedented detail using microcalorimeter
spectrometers.

\section*{Acknowledgements}
\noindent
We thank the \emph{Suzaku} mission managers for executing our requested observations.
We acknowledge helpful conversations with Koji Mukai.

\clearpage

\begin{deluxetable}{l l}
\tablecolumns{2}
\tablewidth{0pc}
\tablecaption{Broken Power-law Model Parameters}
\tablehead{\colhead{Parameter}              &             \colhead{Value}}
\startdata
  	$N_{\mathrm{H}}$($\mathrm{10^{22}}  \mathrm{cm^{-2}}$)	&  		2.00(1)\\	
	$\Gamma_{1}$				&		1.55(1)		\\
	$\Gamma_{2}$				&		2.2(1)		\\
	$E_{\mathrm{break}}$ (keV)				&	36(2)	\\
	$K_{\mathrm{bnpl}}$					&	0.280(3)  \\
 	$F_{0.5-200}$($10^{-9}$)				&			7.51(8)\\
	$L_{\mathrm{X}}$ (0.5$-$200)($\mathrm{10^{37} erg s^{-1}}$) 	&  6.37(7)        \\
	$\chi^{2}$/ $\nu$ 								& 3053/3093 \\
\enddata
\tablecomments{The results of a broadband spectral fit to the \emph{Suzaku} spectra in the 0.7$-$200 keV band are presented above.  The model is described in XSPEC as \texttt{tbabs} x (\texttt{bknpow}).  A constant was allowed to float between the XIS, PIN and GSO data.  The normalization of the broken power-law component is photons $\mathrm{cm^{-2} s^{-1} keV^{-1}}$ at 1 keV. The flux quoted above is an ``unabsorbed'' flux.  The distance to H1743$-$322 is unknown; a Galactic center distance of d = 8.5 kpc was assumed to calculate the luminosity value.  All errors are 90 \% confidence errors.  Errors given in parentheses are symmetric errors in the last significant digit.}
\end{deluxetable}	


\begin{deluxetable}{l l}
\tablecolumns{2}
\tablewidth{0pc}
\tablecaption{Comptonization Model Parameters}
\tablehead{\colhead{Parameter}              &             \colhead{Value}}
\startdata
  	$N_{\mathrm{H}}$($\mathrm{10^{22}} \mathrm{cm^{-2}}$)	&  		1.98(1)\\	
	$T_{\mathrm{o}}$ (keV)				&		0.25(2)		\\
	$T_{\mathrm{e}}$ (keV)				&		38(5)		\\
	$\tau$				&	1.6(2)	\\
	Normalization (x $10^{-2})$					&	1.9(2)   \\
	$\chi^{2}$/ $\nu$ 								& 3066/3092 \\
\enddata
\tablecomments{The results of a broadband spectral fit to the \emph{Suzaku} spectra in the 0.7$-$200 keV band are presented above.  The model is described in XSPEC as \texttt{tbabs} x (\texttt{compTT}).  A constant was allowed to float between the XIS, PIN and GSO data.  The redshift was set to zero and disk geometry was assumed.  Errors given in parentheses are symmetric errors in the last significant digit.}
\end{deluxetable}	



\begin{deluxetable}{c c c c}
\tablecolumns{4}
\tablewidth{0pc}
\tablecaption{X-ray Absorption Lines Upper Limits}
\tablehead{\colhead{Parameter}              &             \colhead{Theoretical Value}			& 		\colhead{Flux}	&		\colhead{W}\\
								&			(\AA)                                               &                     ($10^{-5}$ photons $\mathrm{cm^{-2}}$ $\mathrm{s^{-1}}$)                                                  &                     (m\AA, eV)}
\startdata
	Fe XXV						&			1.850                                            &                       5.1                    &                 0.97, 3.5                     \\
	Fe XXVI						&			1.780                                           &                        3.4                  &                   0.64, 2.5                  \\ 
\enddata
\tablecomments{Parameters for the He-like Fe XXV (1$s^{2}-$1s2p) and H-like Fe XXVI (1s$-$2p) resonance absorption lines.  Since no significant lines were detected, 90\% confidence upper limits are given.}
\end{deluxetable}		


\begin{deluxetable}{c c c c c c}
\tablecolumns{6}
\tablewidth{0pc}
\tablecaption{X-ray Emission Line Detection}
\tablehead{\colhead{Parameter}        &             \colhead{Theoretical Value}	&\colhead{Model Value}  	& \colhead{FWHM} 	& 		\colhead{Flux}	&		                                                                                                                      \colhead{W}\\
					                  	&		     (\AA, keV)                                  & (keV)             					&     (keV)                       &         ($10^{-5}$ photons $\mathrm{cm^{-2}}$ $\mathrm{s^{-1}}$)                                                  &                    (m\AA, eV)}
\startdata
	Low ion. Fe						&	1.940, 6.40                             & 6.3(1)              &          0.6(3)			&	0.3(1)														&	5(2), 17(6)				\\                     
\enddata
\tablecomments{Parameters for low ionization iron (Fe I $-$ Fe XVI), He-like Fe XXV and H-like Fe XXVI  emission lines for the \texttt{tbabs} x (\texttt{gaussian+bknpow}) model.  90\% confidence errors are given.  Low ionization iron with FWHM = 0.6 keV is a weak detection that resulted from allowing the line energy to vary between 6$-$7 keV while $\sigma$ varied between 0 and 3 keV.  Note that the line energy for low ionization iron changes only 0.3 keV between those charge states, which is the reason 6.40 keV was used as the theoretical value.}
\end{deluxetable}

\clearpage

\begin{figure}[t]
\begin{center}
\includegraphics[scale=1.0]{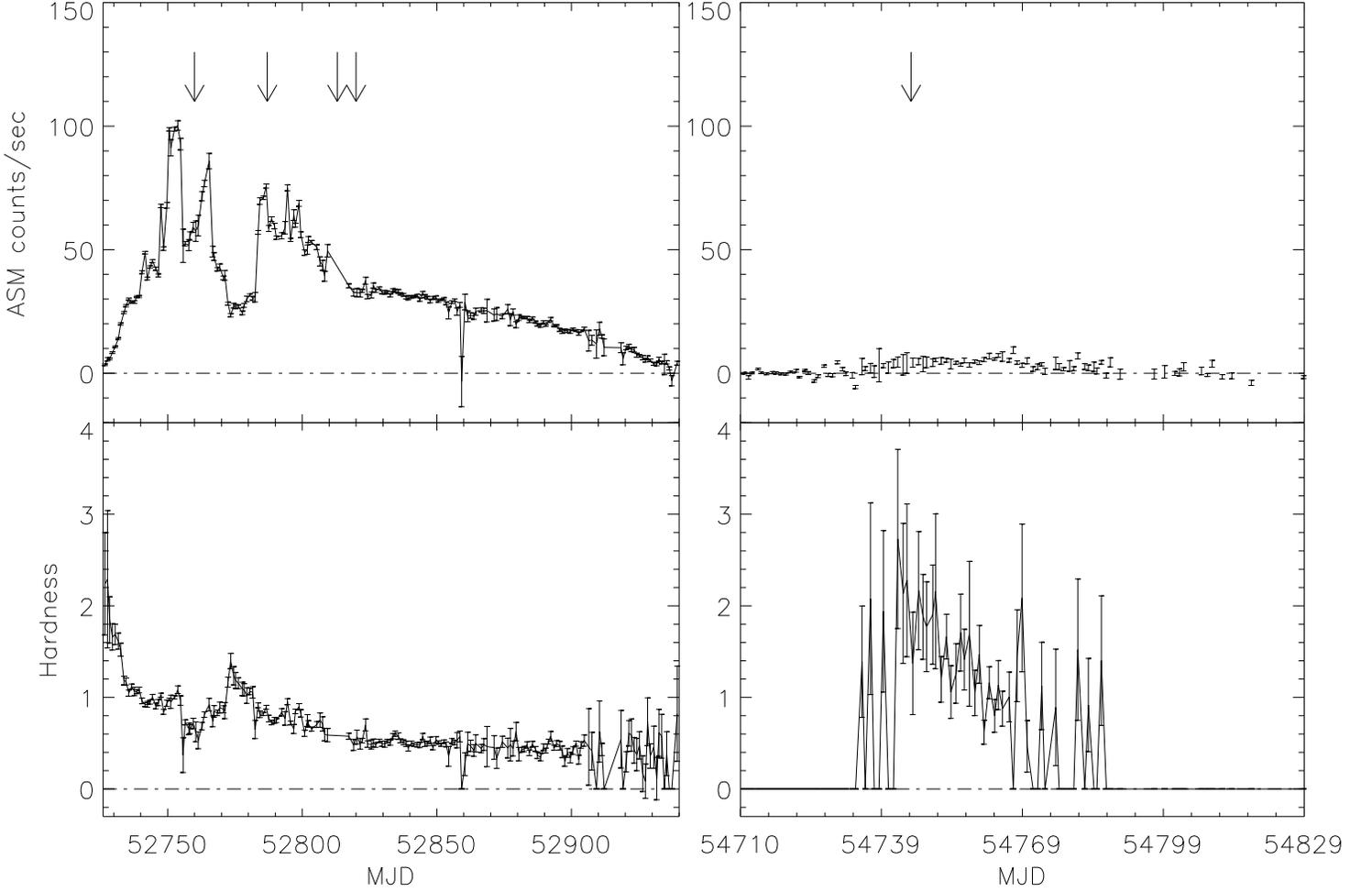}
\vspace{-0.5cm}
\caption{\footnotesize
{\emph{Left panel}: \emph{RXTE/ASM} one-day averaged light curve with errors (1.5$-$12 keV) and one-day averaged (5$-$12 keV)/(3$-$5 keV) hardness ratio, for the bright phase of the 2003 outburst of H1743-322 (Miller et al. 2006c).  The source progresses through the very high or steep power-law state followed by the high/soft state.  The arrows at the top of the plot denote the days on which H1743-322 was observed with \emph{Chandra}.  \emph{Right panel}: \emph{RXTE/ASM} one-day averaged light curve with errors (1.5$-$12 keV) and one-day averaged (5$-$12 keV)/(3$-$5 keV) hardness ratio, for the October 2008 outburst of H1743-322.  The source rise is typical of the low hard state.  The arrow at the top of the plot denotes the date on which we observed H1743-322 with \emph{Suzaku}.}}
\label{fig1}
\end{center}
\end{figure}


\begin{figure}[t]
\begin{center}
\includegraphics[scale=0.60, angle=-90]{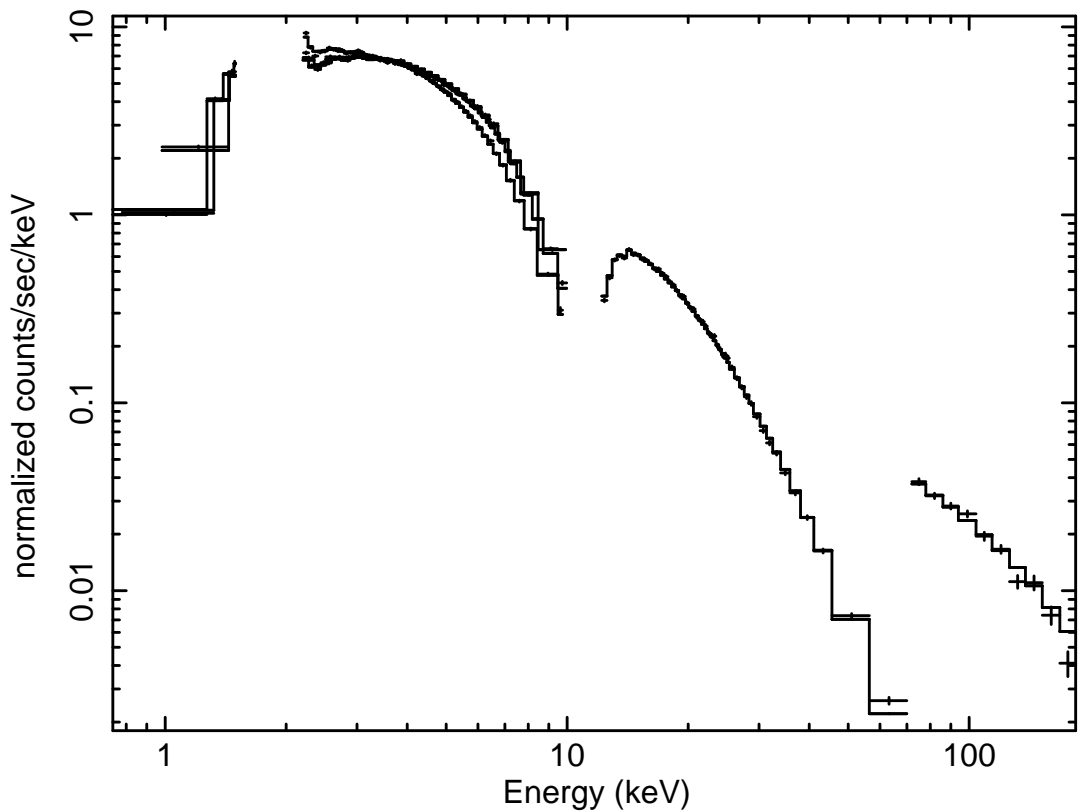}
\vspace{0cm}
\end{center}
\caption{\footnotesize
{\emph{Suzaku} spectrum of H1743-322 fitted with a broken power-law  model (see Table 1).  The curves within the 0.7$-$1.5 keV and 2.2$-$10.0 keV energy ranges represent the data from the XIS detectors.  The curves within the 12.0$-$70.0 keV and 70.0$-$200 keV energy ranges represent the data from the PIN and GSO detectors, respectively.}}
\label{fig2}
\end{figure}


\begin{figure}[t]
\begin{center}
\includegraphics[scale=0.60, angle=-90]{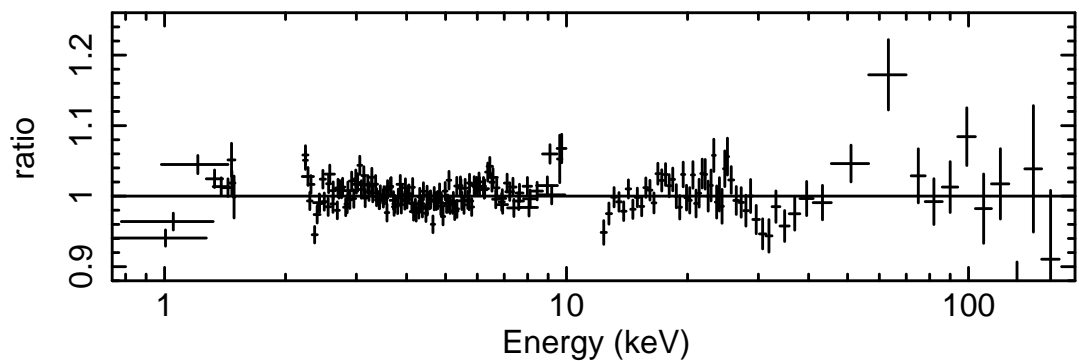}
\vspace{0cm}
\end{center}
\caption{\footnotesize
{Data/model ratio for the broken power-law model.  The curves within the 0.7$-$1.5 keV and 2.2$-$10.0 keV energy ranges represent the data from the XIS detectors.  The curves within the 12.0$-$70.0 keV and 70.0$-$200 keV energy ranges represent the data from the PIN and GSO detectors, respectively.}}
\label{fig3}
\end{figure}


\begin{figure}[t]
\begin{center}
\includegraphics[scale=0.60,angle=-90]{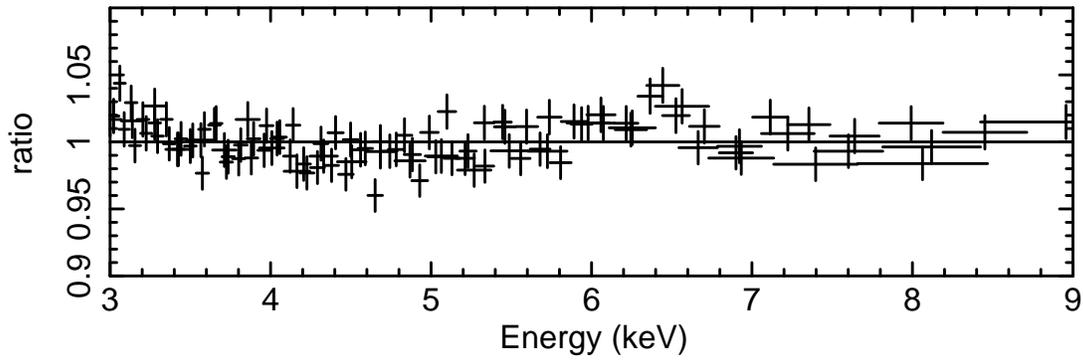}
\vspace{0cm}
\end{center}
\caption{\footnotesize
{Data/model ratio from 3$-$9 keV for the broken power-law model.  The curves represent the data from the XIS detectors.  There is a possible emission line around 6.4 keV.}}
\label{fig4}
\end{figure}


\clearpage

\section*{References}
\noindent
Ballantyne, D. R., Fabian, A. C., \& Ross, R. R. 2002, MNRAS, 329, L67\\
Begelman, M. C., McKee, C. F., \& Shields, G. A. 1983, ApJ, 271, 70\\
Belloni, T., Klein-Wolt, M., M\'endez, M., van der Klis, M., \& van Paradijs, J. 2000, A\&A, 355, 271\\
Blandford, R. D. \& Payne, D. G. 1982, MNRAS, 199, 883\\
Corbel, S., Kaaret, P., Fender, R. P., Tzioumis, A.K., Tomsick, J. A., \& Orosz, J. A. 2005, ApJ, 632, 504\\
Dhawan, V., Mirable, I. F., \& Rodr\'iguez, L. F. 2000, ApJ, 543, 373\\
Doxsey R., Bradt, H., Fabbiano, G., Griffiths, R., Gursky, H., Johnston, M., Leach, R., Ramsey, A., Schwartz, D., Schwarz, J., \& Spada, G. 1977, IAU Circ., 3113, 2\\
Esin, A., McClintock, J., \& Narayan, R. 1997, ApJ, 498, 865\\
Fender, R. \& Belloni, T. 2004, ARA\&A, 42, 317\\
George, I. M., \& Fabian, A. C. 1991, MNRAS, 249, 352\\
Homan, J., Klein-Wolt, M., Rossi, S., Miller, J.M., Wijnands, R., Belloni, T., van der Klis, M., \& Lwein, W. H. G. 2003, ApJ, 586, 1262\\
Homan, K., Miller, J. M., Wijnands, R., van der Klis, M., Belloni, T., Steeghs, D., \& Lewin, W. H. G. 2005, ApJ, 623, 383\\
Kallman, T. R., Palmeri, P., Bautista, M. A., Mendoza, C., \& Krolik, J. H. 2004, ApJS, 155, 675\\
Kaluzienski, L. J. \& Holt, S. S. 1977, IAU Circ., 3099, 3\\
Klein-Wolt, M., Fender, R. P., Pooley, G. G., et al. 2002, MNRAS, 331, 745\\
Koyama, K. et al. 2007, PASJ, 59, S23\\
Laor, A. 1991, ApJ, 376, 90\\
Magdziarz, P., \& Zdziarski, A. A. 1995, MNRAS, 273, 837\\
McClintock, J. E. \& Remillard, R. A. 2006, Compact Stellar X-ray Sources (Cambridge: Cambridge Univ. Press), ed. W. Lewin \& M. van der Klis, 157 - 213\\
Miller, J. M. 2007, ARA\&A, 45, 441\\
Miller, J. M., Homan, J., Steeghs, D., Rupen, M., Hunstead, R. W., Wijnands, R., Charles, P. A., \& Fabian, A. C. 2006a, ApJ, 653, 525\\
Miller, J. M., Raymond, J., Fabian, A., Steeghs, D., Homan, J., Reynolds, C., van der Klis, M., \& Wijnands, R. 2006b, Nature, 441, 953\\
Miller, J. M., Raymond, J., Homan, J., Fabian, A. C., Steeghs, D., Wijnands, R., Rupen, M., Charles, P., van der Klis, M., \& Lewin, W. H. G. 2006c, ApJ, 646, 394\\
Miller, J. M., Raymond, J., Reynolds, C. S., Fabian, A. C., Kallman, T. R., \& Homan, J. 2008, ApJ, 680, 1359\\
Miller, J. M., Wijnands, R., Homan, J., Belloni, T., Pooley, D., Corbel, S., Kouveliotou, C., van der Klis, M., \& Lewin, W. H. G. 2001, ApJ, 563, 928\\
Mitsuda, K., Inoue, H., Koyama, K., Makishima, K., Matsuoka, M., Ogawara, Y., Suzuki, K., Tanaka, Y., Shibazaki, N., \& Hirano, T. 1984, PASJ, 36, 741\\
Neilson, J. \& Lee, J. C. 2009, Nature, 458, 481\\
Ohsuga, K., Mineshige, S., Mori, M., \& Kato, Y. 2009, ArXiv eprints\\
Petrucci, P. O., Merloni, A., Fabian, A., Haardt, F., \& Gallo, E. 2001, MNRAS, 328, 501\\
Proga, D. 2003, ApJ, 585, 406\\
Proga, D., Stone, J. M., \& Kallman, T. R. 2000, ApJ, 543, 686\\
Reis, R. C., Miller, J. M., \& Fabian, A. C. 2009, MNRAS, 395, L52\\
Remillard, R. A. \& McClintock, J. E. 2006, ARA\&A, 44, 49\\
Ross, R. R., Fabian, A. C., \& Young, A. J. 1999, MNRAS, 306, 461\\
Rossi, S., Homan, J., Miller, J. M., \& Belloni, T. 2005, MNRAS 360, 763\\
Rupen, M. P., Mioduszewski, A. J., \& Dhawan, V. 2003, The Astronomer's Telegram, 139\\
Stirling, A. M., Spencer, R. E., de la Force, C. J., Garrett, M. A., Fender, R. P., Ogley, R. N. 2001, MNRAS, 327, 1273\\
Strohmayer, T. E. 2001a, ApJ, 552, L49\\
Strohmayer, T. E. 2001b, ApJ, 554, L169\\
Takahashi, T. et al. 2007, PASJ, 59, 35\\
Titarchuk, L. 1994, ApJ, 434, 570\\
Ueda, Y., Yamaoka, K., \& Remillard, R. 2009, ApJ, 695, 888\\
White, N. E. \& Marshall, F. E. 1984, ApJ, 281, 354\\

\end{document}